\documentclass{mn2e}
\usepackage{graphicx}
\def\go{
\mathrel{\raise.3ex\hbox{$>$}\mkern-14mu\lower0.6ex\hbox{$\sim$}}
}
\def\lo{
\mathrel{\raise.3ex\hbox{$<$}\mkern-14mu\lower0.6ex\hbox{$\sim$}}
}
\def\simeq{
\mathrel{\raise.3ex\hbox{$\sim$}\mkern-14mu\lower0.4ex\hbox{$-$}}
}

\def\etal{{\it et al.\ }}

\def\etal{{\it et al.\ }}

\def\be{\begin{equation}}
\def\ee{\end{equation}}
\def\bea{\begin{eqnarray}}
\def\eea{\end{eqnarray}}

\def\etal{{\sl et al.\ }}

\def\hw2{{\hat W}^2}
\def\go{\mathrel{\raise.3ex\hbox{$>$}\mkern-14mu
             \lower0.6ex\hbox{$\sim$}}}
\def\lo{\mathrel{\raise.3ex\hbox{$<$}\mkern-14mu
             \lower0.6ex\hbox{$\sim$}}}
\def\ltorder{\mathrel{\raise.3ex\hbox{$<$}\mkern-14mu
             \lower0.6ex\hbox{$\sim$}}}
\def\gtorder{\mathrel{\raise.3ex\hbox{$>$}\mkern-14mu
             \lower0.6ex\hbox{$\sim$}}}

\def\eps2{{\epsilon^2}}

\begin{document}

\title
[The connection between Ultraviolet and X-ray Outflows]
{The Connection between Ultraviolet and X-ray Outflows in AGN: the case
of PDS~456}
\author[Paul T. O'Brien  \etal]{Paul T. O'Brien$^{1}$, James N. Reeves$^{2,3}$,
Chris Simpson$^{4}$, Martin J. Ward$^{1,4}$\\
$^{1}$ X-Ray and Observational Astronomy Group, 
Department of Physics \& Astronomy,  
University of Leicester, Leicester, LE1 7RH, UK\\
$^{2}$ Laboratory for High Energy Astrophysics, Code 662, NASA Goddard
Space Flight Center, Greenbelt Road, Greenbelt, MD20771, USA\\
$^{3}$ Department of Physics and Astronomy, 
Johns Hopkins University, 3400 N Charles Street, Baltimore, MD 21218,
USA\\
$^{4}$ Department of Physics, University of Durham, Rochester
Building, Science Laboratories, South Road, Durham, DH1 3LE
}

\date{Received ** *** 2004 / Accepted ** *** 2004}

\label{firstpage}

\maketitle

\begin{abstract}
High-velocity outflows from AGN are a well-known phenomena in terms of
the Broad Absorption Lines seen in the UV/optical. More recently,
similar, possibly related, outflows have been reported in the X-ray.
The most extreme example is seen in the nearby, luminous QSO PDS~456,
which displays a massive, high velocity (50000 km s$^{-1}$),
high-ionization X-ray outflow of $10\, {\rm M}_{\odot}$ yr$^{-1}$.
Here we present the UV spectrum of PDS~456 as observed by the Hubble
Space Telescope. We find the UV spectrum is also extreme, displaying
very broad emission-lines, with C~{\sc iv} $\lambda 1549$ blueshifted
by 5000 km s$^{-1}$ and a broad Ly$\alpha$ absorption trough
blueshifted by 14000--24000 km s$^{-1}$. No strong, broad
high-ionization absorption features are seen. We interpret the
combined UV and X-ray spectrum of PDS~456 as the signature of a
decelerating, cooling outflow, which may be driven by radiation and/or
magnetic field. This outflow may be the source of some of the broad
emission and absorption-line gas.

\end{abstract}

\begin{keywords}
galaxies: active -- UV: galaxies -- galaxies: individual: PDS~456
\end{keywords}

\section{Introduction}
\label{sec:intro}

If we are to understand the nature of Active Galactic Nuclei (AGN), it
is crucial to fully explore the available range of parameter space and
in particular study objects which display the most extreme behaviour.
One such object is the nearby quasar PDS~456, which was discovered by
Torres et al.\ (1997). They suggested it was intrinsically more
luminous than 3C~273, traditionally taken to be the most luminous
object in the ``local Universe'' ($z < 0.3$). Our follow-up
observations confirmed that PDS~456 is more luminous than 3C~273 in
the optical, with $M_B = -26.7$ (Simpson et al.\ 1999). The reason
PDS~456 was not discovered sooner is due to its sky location, near the
Galactic bulge, resulting in Galactic extinction of E(B$-$V) = 0.48.

Unlike 3C~273, VLA radio data show PDS~456 to be radio-quiet with a
typical optical-to-radio luminosity ratio for that class
(Simpson et al. 1999), suggesting its high luminosity is not due to
Doppler boosting of the continuum. In the near-infrared and optical
PDS~456 is effectively a ``twin'' of 3C~273 in terms of its continuum
shape, and displays a rich emission-line spectrum in which the Balmer
and Paschen lines have very broad wings ({\rm FWZI} $>
30000$\,km\,s$^{-1}$) and the Fe~{\sc ii} lines are strong. The
[O~{\sc iii}]~$\lambda$5007 line is very weak (EW $<
2$\,\AA, compared to a typical value of 24\,\AA; Miller et al.\ 1992).
Simpson et al. derive a redshift from the narrow [Fe~{\sc ii}]
$\lambda 1.6435$ $\mu$m line of $0.18375 \pm 0.00030$. 

Using ASCA and RXTE data, Reeves et al. (2000) showed that
PDS~456 has a steep hard X-ray spectrum (photon index $\Gamma \sim
2$), with an unusually strong absorption edge at $\approx 8.9$\,keV
(in the rest-frame). A higher-quality observation obtained with
XMM-Newton confirms the rich, absorption-dominated X-ray spectrum
(Reeves et al. 2002; Reeves, O'Brien \& Ward 2003). To fit these
data requires the presence of a large column ($N_H = 5
\times 10^{23}$ cm$^{-2}$) of highly ionized material (log $\xi =
2.5$). This material is outflowing at $\sim 50000$ km s$^{-1}$, which
translates to a mass of $10 \, {\rm M}_{\odot}$ yr$^{-1}$ for a
conservative covering factor of 0.1 steradian. Similar, but less
extreme, high-velocity outflows have also been observed in other AGN,
(Chartas et al. 2002; Chartas, Brandt \& Gallagher 2003; Pounds et al.
2003a,b; Dasgupta et al. 2005). An alternative viewpoint has been
expressed that some of the absorption could be due to local hot gas
(e.g. McKernan, Yaqoob \& Reynolds 2004). While this may explain a
modest column of soft X-ray absorbing gas, such a model does not
provide a good fit to the strong Fe K-band features seen in PDS~456
and has no explanation in terms of the known local Galactic
environment.

Correcting for the Galactic and intrinsic absorption, PDS~456 has an
X-ray luminosity of L(2--10 keV) $\approx 8.1 \times
10^{44}$\,erg\,s$^{-1}$ (adopting H$_0 = 75$ km s$^{-1}$ Mpc$^{-1}$,
$\Omega_{\rm M} = 0.3$, $\Omega_{\Lambda}=0.7$). This is about a tenth
of the X-ray luminosity of 3C~273, but PDS~456 has an optical/X-ray
luminosity ratio within the known range for radio-quiet AGN (e.g. Yuan
et al. 1998). The X-ray variability of PDS~456 is extreme for such a
luminous radio-quiet object. It varies by factors of two in just 30
ks, repeating this behaviour almost every day (Reeves et al. 2002).
Overall, the luminosity, variability and spectral characteristics of
PDS~456 strongly suggest it is an object running at an unusually high
accretion rate, with a massive X-ray outflow driven by radiation
and/or magnetic field.

The X-ray properties of PDS~456 clearly show it to be an extraordinary
object. To explore its nature further, we report here on a
HST UV observation which reveal a further set of extreme properties.
In the following sections we describe the HST observation of PDS~456
and the derived spectral characteristics. We then discuss how the UV
data suggest a connection between the X-ray and UV properties and how
PDS~456 could help us understand the connection between AGN with and
without broad absorption lines and explain the origin of at least part
of the broad emission line region.

\section{The HST observation}

We observed PDS~456 with the HST Space Telescope Imaging Spectrograph
(STIS) on 2000 May 14. Exposure times of 1140 and 781 seconds were
used for the G140L and G230L gratings respectively. Galactic reddening
results in an {\it observed} flux level of only 2--10 $\times
10^{-15}$ erg cm$^{-2}$ s$^{-1}$ in the STIS data, but nevertheless
PDS~456 is clearly detected throughout the entire wavelength range.
The dereddened UV spectrum is shown in Fig. 1. An E(B$-$V) = 0.48 was
adopted (Torres et al. 1997; Simpson et al. 1999) with the Seaton
(1979) extinction law. As this was an exploratory observation using a
single HST orbit, the signal-to-noise ratio is not very high, but
nevertheless several broad emission features are detected with high
significance (7--10$\sigma$), well above any known calibration or
instrumental uncertainties. Many narrow Galactic absorption lines are
also detected, but will not be discussed further in this paper. The
dereddened spectra were combined and analysed using the Starlink~{\sc
DIPSO} software program (Howarth et al. 1998).

\section{The UV properties of PDS~456}

\begin{figure}
\centering
\includegraphics[clip,width=1.0\linewidth,angle=0]{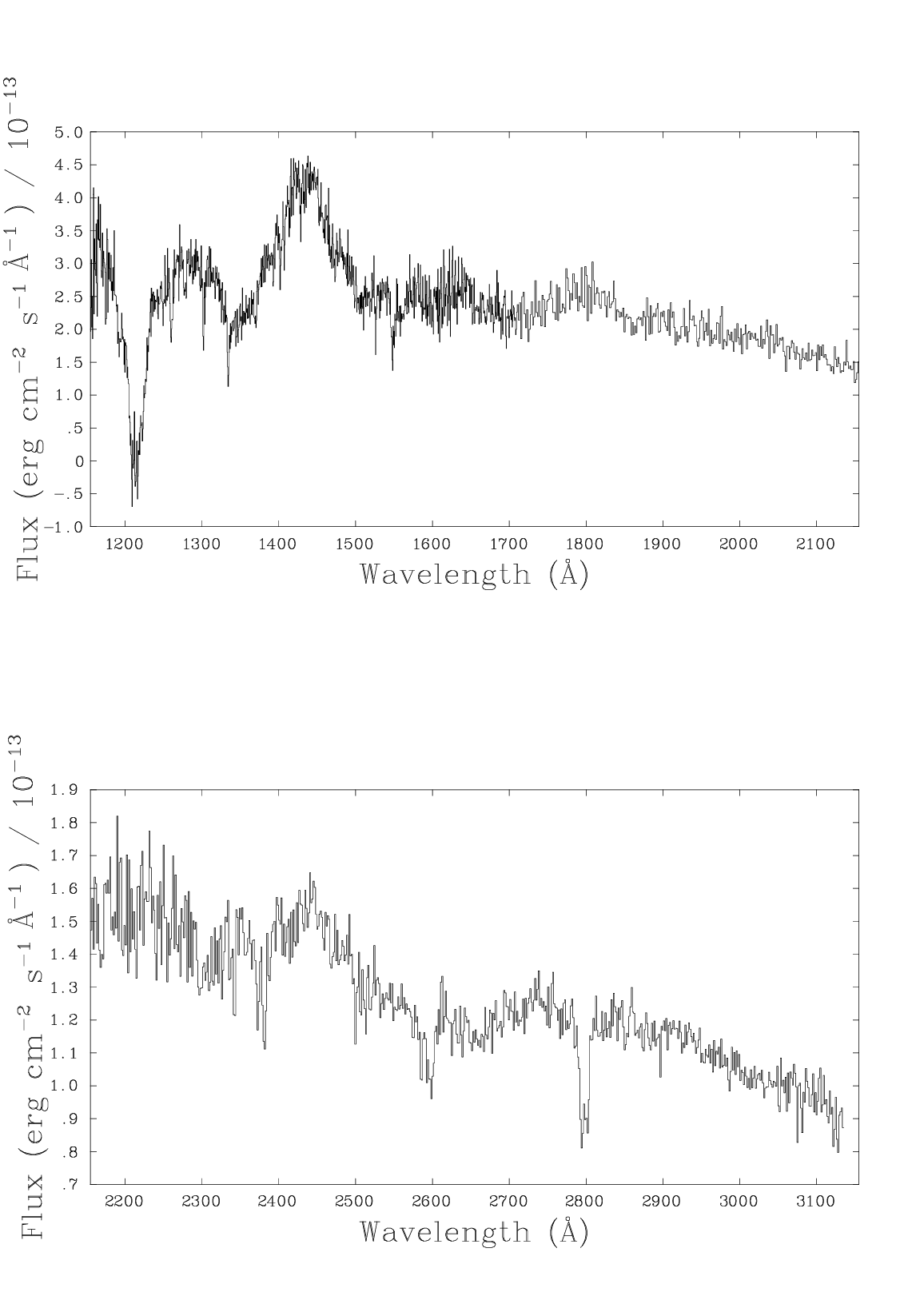}
\caption[]{The observed UV spectrum of PDS~456 corrected for Galactic
reddening with E(B$-$V) = 0.48. Several emission and absorption
features can be seen, including Ly$\alpha$/N~{\sc v} $\lambda 1240$
(at $\approx 1435$\AA) and blueshifted Si {\sc iv}/O~{\sc iv} $\lambda
1400$ (at $\approx 1635$\AA ) and C~{\sc iv} $\lambda 1549$ (at
$\approx 1800$\AA ). An intrinsic broad absorption feature, ascribed
to Ly$\alpha$, is also present at $\approx 1350$\AA . Numerous narrow
Galactic absorption features are also detected, including Si~{\sc ii}
$\lambda 1260$, O~{\sc i} $\lambda 1302$, C~{\sc ii} 1335, Si~{\sc
iv} $\lambda\lambda 1394, 1403$, C~{\sc iv} $\lambda\lambda 1548,
1551$, Mg~{\sc ii} $\lambda\lambda 2796, 2803$, and several Fe~{\sc
ii} lines.}
\end{figure}

\begin{figure}
\centering
\includegraphics[clip,width=0.7\linewidth,angle=-90]{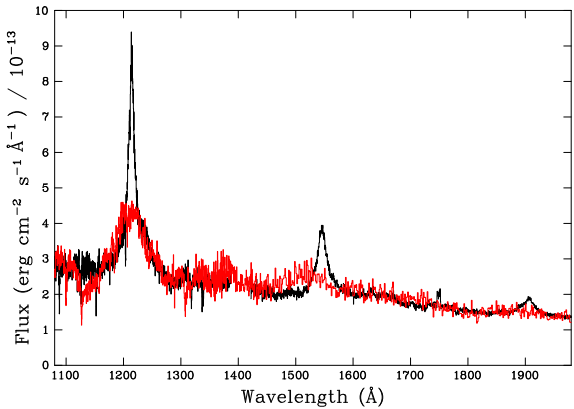}
\caption[]{The dereddened STIS UV spectra of PDS~456
compared to an archival HST/FOS UV spectrum of 3C~273 (dark grey
line). Both wavelength ranges have been converted to the rest-frame.
The UV continuum shape and strength is remarkably similar in both
objects, while the emission and absorption features are strikingly
different. The Ly$\alpha$ emission line in PDS~456 lacks the strong
core seen in 3C~273, the C~{\sc iv} $\lambda 1549$ emission line is
strongly blueshifted in PDS~456 and again is less peaked, the C~{\sc
iii}] $\lambda 1909$ emission line is not detected in PDS~456, and
PDS~456 has a strong, broad absorption feature blueward of Ly$\alpha$
which is not seen in 3C~273.}
\end{figure}

\begin{figure}
\centering
\includegraphics[clip,width=0.7\linewidth,angle=-90]{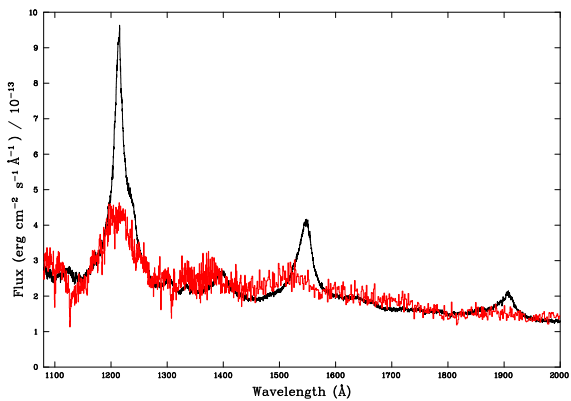}
\caption[]{The dereddened STIS UV spectra of PDS~456
compared to the mean HST/FOS AGN spectrum from Telfer et al. (2002)
(light grey line). Both wavelength ranges have been converted to the
rest-frame.}
\end{figure}


\begin{table*}
\begin{center}
\caption{UV emission-line data for PDS~456.} 
\label{lines}
$^{a}$ Observed wavelength.
$^{b}$ Observed Line flux.
$^{c}$ Velocity shift relative to $z=0.18375$.
\begin{tabular}{@{}lcccc}
\hline
Line & $\lambda^a_{obs}$ & Flux$^b$ & FWHM & $\Delta v^c$ \\ 
 & (\AA) & ($10^{-12}$ erg cm$^{-2}$ s$^{-1}$) & (km s$^{-1}$) & (km
s$^{-1}$)\\
\hline
Ly$\alpha$/N~{\sc v} $\lambda 1240$ (1 Gaussian) & $1436.2 \pm 0.6$ &
$11.6\pm0.25$ & $13578\pm334$ & $\ -594\pm125$ \\
Ly$\alpha$/N~{\sc v} $\lambda 1240$ (2 Gaussian) & $1429.7 \pm 2.3$ &
$8.6\pm0.9$ & $11961\pm567$ & $-1949\pm479$ \\
 & $1458.7\pm2.3$ & $2.9\pm0.9$ & $11961\pm567$ & $-1949\pm479$ \\
Si~{\sc IV}/O~{\sc iv} $\lambda 1400$ & $1635.4\pm2.4$ & $1.4\pm0.2$ &
$\ \ \ 6879\pm1229$ & $-3955\pm434$ \\
C~{\sc iv} $\lambda 1549$ & $1801.6\pm3.3$ & $5.1\pm0.4$ & $\ 14787\pm1415$ &
$-5240\pm539$ \\
\hline
\end{tabular}
\end{center}
\end{table*}

The continuum shape in PDS~456 is fairly ``normal'' with a UV spectral
index $\alpha \approx -0.7$ ($f_{\nu} \propto \nu^{\alpha}$), agreeing
with an extrapolation of our previous optical spectra (from August
1997) and with the XMM-Newton Optical Monitor broad-band UV fluxes
(from May 2002), suggesting little continuum variability occurs in the
UV. Simpson et al. (1999) noted that the optical continuum shape of
PDS~456 is very similar to that of 3C~273, and this similarity extends
into the UV band, as shown in Fig. 2. The UV continuum shape of
PDS~456 is also similar to that of the mean radio-quiet quasar UV
spectrum compiled by Telfer et al. (2002) from HST spectra (Fig. 3).

To extract spectral information we fitted the ``expected'' emission
lines with Gaussian profiles, trying both a single and double-Gaussian
fit to the Ly$\alpha$/N~{\sc v} $\lambda 1240$ complex. The derived
parameters are given in Table 1.

The UV emission lines from the broad-line region (BLR) in PDS~456 are
unusually broad (FWHM up to $\sim 14000$ km s$^{-1}$). Simpson et al.
(1999) note that the optical/IR hydrogen lines have a very broad base
(FWZI $> 30000$ km s$^{-1}$), and the UV spectrum shows these are
similar in shape to Ly$\alpha$/N~{\sc v} $\lambda 1240$ (Fig. 4). The
Ly$\alpha$/N~{\sc v} $\lambda1240$ profile in PDS~456 is also similar
to the base of this feature in 3C~273 (Fig. 2). The broad UV emission
lines in PDS~456 lack a sharp, low-velocity core unlike those seen in
the Balmer lines (Fig. 4) or in the UV lines of 3C~273 and the mean
HST spectrum.

Broad C~{\sc iv} $\lambda 1549$ and Si~{\sc iv}/O~{\sc iv} $\lambda
1400$ lines are detected in PDS~456, but are both strongly blueshifted
by 5000 and 4000 km s$^{-1}$ respectively relative to the optical/IR
low-ionization lines. The blueshift of 2000 km s$^{-1}$ for
Ly$\alpha$/N~{\sc v} $\lambda 1240$ is significantly smaller,
consistent with the trend in blueshift found by Vanden Berk et al.
(2001). The double-Gaussian fit given in Table~1 for Ly$\alpha$/N~{\sc
v} $\lambda 1240$ was performed assuming the two lines to have the
same width and velocity shift. Allowing the line centres to go free
gave a consistent but less well constrained fit.

The absence of a sharp low-velocity core in
Ly$\alpha$/N~{\sc v} $\lambda 1240$ and C~{\sc iv} $\lambda 1549$
results in both these features having about half the rest-frame
equivalent widths (38 and 22\AA\ respectively) of those in 3C~273 or
in the mean HST spectrum. Despite the blueshifts, the
Ly$\alpha$/N~{\sc v} $\lambda 1240$ and C~{\sc iv} profiles are quite
similar (Fig. 5). Neither the He~{\sc ii} $\lambda 1640$ nor C~{\sc
iii}] $\lambda 1909$ lines are clearly detected, the latter possibly
indicating a dense BLR. The upper limit for C~{\sc iii}] ($1.2 \times
10^{-12}$ erg cm$^{-2}$ s$^{-1}$ gives a line strength relative to
C~{\sc iv} of about half the average (Kuraszkiewicz et al. 2002;
Telfer et al. 2002).

Aside from the blueshifted C~{\sc iv} $\lambda 1549$ line, the most
striking feature in the UV spectrum of PDS~456 is a broad absorption
trough centred around an observed wavelength of $\lambda 1350$. For
illustration, we show in Fig. 6 the observed spectrum of PDS~456
shortward of the Ly$\alpha$/N~{\sc v} $\lambda 1240$ emission line
compared to that of a typical, low-redshift AGN, NGC~3783. The same
narrow, Galactic absorption features can be seen, but not the broad
absorption trough. If identified as Ly$\alpha$ absorption, the broad
absorption feature in PDS~456 is blueshifted by 14000--24000 km
s$^{-1}$ relative to the optical/IR narrow emission lines.

\section{Discussion}

The UV spectrum of PDS~456 displays three unusual properties: very
large emission lines blueshifts of 4000--5000 km s$^{-1}$ for the C~{\sc
iv} $\lambda 1549$ and Si~{\sc IV}/O~{\sc iv} $\lambda 1400$ lines;
very broad UV emission lines (FWHM $\approx 14000$ km s$^{-1}$) which
lack a narrow core; and a broad, blueshifted absorption trough, which
if identified with Ly$\alpha$ extends from $-14000$ to $-24000$ km
s$^{-1}$. The fact that these features are seen in a quasar with a
massive, high velocity, high-ionization X-ray outflow strongly
suggests a connection between the UV and X-ray properties.

\subsection{Emission Line Properties}

Comparing with the UV emission-line properties of AGN derived by
Kuraszkiewicz et al. (2002), the broad-line widths for PDS~456 are in
the upper 10 per-cent of the population for Ly$\alpha$/N~{\sc v}
$\lambda 1240$ and C~{\sc iv} $\lambda 1549$. If we compare with the
broad base of the lines, as fitted by Kuraszkiewicz et al., the line
widths in PDS~456 are still in the upper 20\% of the population. The
line strengths relative to the UV continuum are about half the mean
values except for Si~{\sc iv}/O~{\sc iv}. The ratios of the higher
ionization broad lines to Ly$\alpha$ in PDS~456 are at the high end of
the observed distribution for AGN.

The biggest difference in broad line properties for PDS~456 comes when
comparing the blueshifts. It is well known that high-ionization broad
emission-lines are often blueshifted relative to the low-ionization
lines (e.g. Wilkes 1984, Espey et al. 1989), but the blueshifts are
usually modest. For 794 radio-quiet quasars Richards et al. (2002)
find a median C~{\sc iv} $\lambda 1549$ blueshift relative to Mg~{\sc
ii} $\lambda 2798$ of 824 km s$^{-1}$. They find blueshifts above 2000
km s$^{-1}$ are rare and none of their sample have confirmed values
above 3000 km s$^{-1}$. At 5000 km s$^{-1}$, PDS~456 has the largest
C~{\sc iv} blueshift known to date.

To explain the C~{\sc iv} $\lambda 1549$ emission-line blueshifts,
Richards et al. (2002) propose an absorption model, in which objects
with larger blueshifts are those in which the red wing of the
emission-line has been absorbed. We note that PDS~456 has a C~{\sc iv}
blueshift about double that of any object in the Richards et al.
sample. Sub-dividing their sample by increasing amounts of blueshift,
Richards et al. find those with larger blueshifts have a weaker than
average Ly$\alpha$/N~{\sc v} $\lambda 1240$ ratio and weaker He~{\sc
ii} $\lambda 1640$, but a similar Si~{\sc iv}/O~{\sc iv} $\lambda
1400$ ratio. Although at first sight PDS~456 does not seem to violate
these trends, if we wish to explain the blueshift by saying C~{\sc iv}
is strongly absorbed in the red wing it would imply a very large
amount of absorption, an extremely large velocity range for the
``unabsorbed'' line and would not explain why it looks similar in
profile to Ly$\alpha$/N~{\sc v} $\lambda 1240$ and a similarly
blueshifted, albeit noisy, Si~{\sc iv}/O~{\sc iv} feature.

\begin{figure}
\centering
\includegraphics[clip,width=0.7\linewidth,angle=-90]{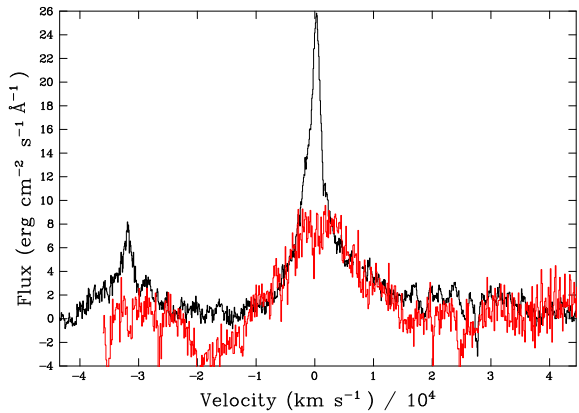}
\caption[]{A comparison between the scaled profiles of the Ly$\alpha$/N~{\sc v}
and H$\beta$ (dark grey line) emission-lines with a locally fitted
continuum subtracted from both. Fe~{\sc ii} emission has been subtracted from
the optical spectrum (Simpson et al. 1999)).}
\end{figure}

\begin{figure}
\centering
\includegraphics[clip,width=0.7\linewidth,angle=-90]{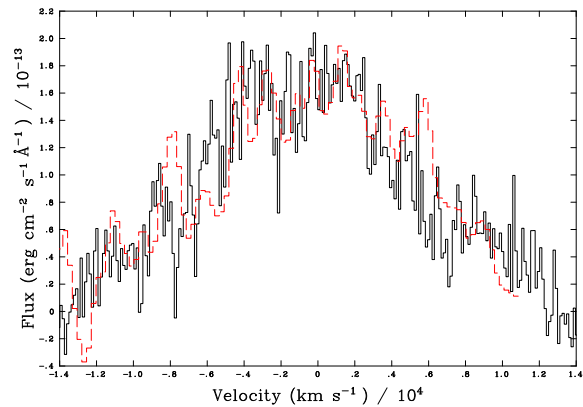}
\caption[]{A comparison between the profiles of the Ly$\alpha$/N~{\sc v}
(solid) and C~{\sc iv} $\lambda 1549$ (dashed) emission-lines with a
locally fitted continuum subtracted from both. The C~{\sc iv} profile
has been scaled and slightly smoothed.}
\end{figure}

\begin{figure}
\centering
\includegraphics[clip,width=0.7\linewidth,angle=-90]{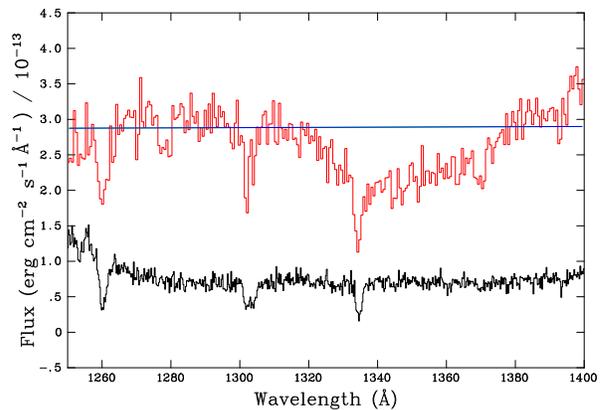}
\caption[]{A comparison between the rest-frame UV spectra of PDS~456
(upper) and that of the Seyfert 1 galaxy NGC~3783 (lower). Narrow,
Galactic absorption features due to Si~{\sc ii} $\lambda 1260$, O~{\sc
i} $\lambda 1302$ and C~{\sc ii} 1335 are clearly seen in both
spectra whereas the broad absorption is only seen in PDS~456. The
horizontal line drawn on the PDS~456 spectrum is merely to guide the
eye.}
\end{figure}

\subsection{Outflows and the Broad Emission Lines}

Broad UV absorption features of similar strength and velocity width to
that of the Ly$\alpha$ absorption feature seen in PDS~456 are seen in
broad absorption-line quasars (BALs; e.g. Turnshek 1984). However, we
do not unambiguously detect the usual N~{\sc v} $\lambda 1240$ and
C~{\sc iv} $\lambda 1549$ BALs. Unfortunately the Mg~{\sc ii} $\lambda
2798$ feature falls at the atmospheric cut-off of the optical spectrum
and is beyond the range of the HST spectrum. Hence we cannot place
firm constraints on the ionization state of the UV absorber. It could
be argued that the UV broad absorption feature in PDS~456 is actually
highly blueshifted N~{\sc v} $\lambda 1240$ absorption, but then we
cannot explain the lack of strong C~{\sc iv} $\lambda 1549$ absorption
as both lines are commonly seen, with similar strength, in BAL
quasars. Similarly, if we assume the lack of a narrow emission-line
core in Ly$\alpha$ is due to N~{\sc v} $\lambda 1240$ absorption, this
would place the N~{\sc v} BAL at a much lower velocity than the
Ly$\alpha$ absorption. The detection of a single BAL feature poses a
problem in comparing with traditional BAL quasars and it could be
argued PDS~456 is a different sort of AGN. Nevertheless, we discuss
PDS~456 below in the context of such objects.

The origin of the BAL phenomenon in QSOs is unclear, but recent
studies suggest they are drawn from the same parent population as
other quasars (Reichard et al. 2003). If so, that implies BALs are
present in all AGN but are only visible from certain orientations
(e.g. Weymann et al. 1991). Many BAL acceleration mechanisms have been
proposed, such as line-driving (Murray et al. 1995) or hydromagnetic
driving (K\"onigl \& Kartje 1994). Line-driven winds models are the
most commonly considered. The observed ionization state of
the UV absorbing gas implies it must be shielded from the full
ionizing continuum (Murray et al.) and such gas arises naturally in
simulations of line-driven accretion disk winds (e.g. Proga, Stone
\& Kallman 2000). The shielding gas has sufficient column density to
provide significant soft X-ray absorption in BAL quasars, as appears
to be the case for some objects (e.g., Gallagher et al. 2002).

In the context of BAL wind models, PDS~456 could be viewed as a
high-ionization analogue of the BAL quasars, in which the bulk of the
absorbing gas, including the ``shield'', has been ionized, and is
observed as a high ionization, high-velocity absorber in the X-ray
(Reeves, O'Brien \& Ward 2003). The implied high accretion rate and
large C~{\sc iv} blueshift in PDS~456 are properties associated with
an increased likelyhood of absorption/BALs (Reichard et al. 2003;
Gallagher et al. 2004). However, this apparent similarity should not
be taken as proof that the X-ray outflow gas in PDS~456 is simply
driven in the same way as a BAL. It is hard to envisage a scenario in
which a highly-ionized, high velocity outflow such as that in PDS~456
is driven purely by line or continuum radiation (see also Everett \&
Ballantyne 2004). The evidence for magnetic flaring in PDS~456 (Reeves
et al. 2002) suggest rather that hydromagnetic driving could be an
important driving mechanism. If so, the geometry of the outflow need
not be along the disk plane, as commonly invoked for accretion disk
wind models. We note that the kinetic energy of the X-ray outflow
($\sim 10^{46}$ erg s$^{-1}$) is comparable to the energy input
required to explain the largest radio galaxies, assuming they are
built over $\sim 10^8$ yr. The outflow could, therefore, be viewed as
a wider, slower, higher-mass version of a relativistic jet.

PDS~456 has a far-infrared luminosity comparable to that of an
Ultra-Luminous Infrared Galaxy (ULIRG), such as Arp~220, but it has a
relatively low CO luminosity (Yun et al. 2004), and hence a relatively
low mass of molecular gas. Yun et al. suggest that it may be evolving
from being a ULIRG into a QSO. The massive, highly-ionized outflow in
PDS~456, indicative of a massive black hole accreting at a high rate,
may then be due to it having undergone a recent ``accretion event''
which has triggered a phase of high activity. PDS~456 may therefore be
a relatively young, or refuelled, AGN, which would help explain it
having a bolometric luminosity more typical of $z\sim3$ objects.

The large outflow velocity found for the X-ray absorber (blueshift
$\sim 50000$ km s$^{-1}$) is much larger than that of the UV broad
absorption (blueshift 14000--24000 km s$^{-1}$) and the broad emission
lines (blueshifts up to 5000 km s$^{-1}$). One possibility is that the
UV features are due to a decelerating, cooling outflow. In this
scenario the X-ray outflow could be the source of at least some of the
BLR, most likely the high-ionization gas component. The mass-loss rate
derived from the X-ray data of $10\, {\rm M}_{\odot}$ yr$^{-1}$ is
more than sufficient to replenish the $\le 1$ M$_{\odot}$ of gas
required to explain the observed C~{\sc iv} $\lambda$ 1549 emission.
To constrain this model we need to place firm limits on the N~{\sc v},
Si~{\sc iv} and C~{\sc iv} absorption features which are commonly seen
in BAL quasars. If no UV high-ionization absorption were detected but
only Ly$\alpha$, that could imply the Ly$\alpha$ absorber is a
different component, which would be curious given its similar
blueshift to the highly-ionized X-ray absorbing material. To obtain a
higher-quality UV spectrum requires the repair of HST or the provision
of a replacement. Also crucial to constraining the otuflow will be a
recently approved higher-resolution Astro-E2/XRS observation. These data
will probe the velocity and ionization structure and possibly reveal
lower ionization X-ray features analagous to those seen in the UV.

\section*{Acknowledgements}
Based on observations made with the NASA/ESA Hubble Space Telescope,
obtained at the Space Telescope Science Institute, which is operated
by the Association of Universities for Research in Astronomy, Inc.,
under NASA contract NAS 5-26555. These observations are associated
with program 8264.

\section*{References}

Chartas, G., Brandt, W.N., Gallagher, S.C., Garmire, G.P., 2002, ApJ,
579, 169\\
Chartas, G., Brandt, W.N., Gallagher, S.C., 2003, ApJ, 595, 85\\
Dasgupta, S., Rao, A.R., Dewangan, G.C., Agrawal, V.K., 2005, ApJ,
618, L87\\
Espey, B.R., Carswell, R.F., Bailey, J.A., Smith, M.G., Ward, M.J.,
1989, ApJ, 342, 666\\
Everett, J.E., Ballantyne, D.R., 2004, ApJ, 615, L13\\
Gallagher, S.C., Brandt, W.N., Chartas, G., Garmine, G.P., 2002, ApJ,
567, 37\\
Gallagher, S.C., Richards, G.T., Hall, P.B., Brandt, W.N., Schneider,
D.P., Vanden Berk, D.E., 2004, ApJ, in press (astro-ph/0410641)\\
Howarth, I.D., Murray, J., Mills, D., Berry, D.S., Starlink User Note
50.21, Rutherford Appleton Laboratory \\
K\"onigl, A., Kartje, J.F., 1994, ApJ, 424, 446\\
Kuraszkiewicz, J.A., Green, P.J., Forster, K., Aldcroft, T.L., Evans,
I.N., Koratkar, A., 2002, ApJS, 143, 257\\
McKernan, B., Yaqoob, T., Reynolds, C.S., 2004, ApJ, 617, 232\\
Miller P., Rawlings S., Saunders S., Eales S., 1992, MNRAS, 254, 93 \\
Murray, N., Chiang, J., Grossman, S.A., Voit, G.M., 1995, ApJ, 451,
498\\
Pounds, K.A., Reeves, J.N., King, A.R., Page, K.L., O'Brien, P.T.,
Turner, M.J.L., 2003a, MNRAS, 345, 705\\
Pounds, K.A., King, A.R., Page, K.L., O'Brien, P.T., 2003b, MNRAS, 346, 1025\\
Proga, D., Stone, J.M., Kallman, T.R., 2000, ApJ, 543, 686\\
Reeves, J.N., O'Brien, P.T., Vaughan, S., Law-Green, D., Ward, M.J.,
Simpson, C., Pounds, K.A., Edelson, R.A., 2000, MNRAS, 312, L17 \\
Reeves, J.N. Wynn, G., O'Brien P.T., Pounds, K.A., 2002, MNRAS, 336,
L56 \\
Reeves, J.N., O'Brien, P.T., Ward, M.J., 2003, ApJ, 593, L65\\
Reichard, T.A., et al., 2003, AJ, 126, 259\\
Richards, G.T., Vanden Berk, D.E., Reichard, T.A., Hall, P.B.,
Schneider, D.P., Subbarao, M., Thakar, A.R., York, D.G., 2002, AJ,
124, 1\\
Seaton, M., 1979, MNRAS, 187, 73P\\
Simpson C., Ward M.J., O'Brien P.T., Reeves J.N., 1999, MNRAS,
303, L23\\
Telfer, R.C., Zheng, W. Kriss, G.A., Davidsen, A.F., 2002, ApJ, 565,
773\\
Torres C.A.O., et al., 1997, ApJ, 488, L19 \\
Turnshek, D.A., 1984, ApJ, 280, 51 \\
Vanden Berk et al., 2001, AJ, 122, 549\\
Weymann, R.J., Morris, S.L., Foltz, C.B., Hewett, P.C., 1991, ApJ,
373, 23\\
Wilkes, B.J., 1984, MNRAS, 207, 73\\
Yuan, W., Brinkmann, W., Siebert, J., Voges, W., 1998, A\&A, 330, 108\\
Yun, M.S.,  Reddy, N.A., Scoville, N.Z., Frayer, D.T., Robson, E.I., 
Tilanus, R.P.J., 2004, ApJ, 601, 723\\

\end{document}